\documentclass[10pt,superscriptaddress,twocolumn,amsmath,amssymb,aps,prl,reprint]{revtex4-1}

\usepackage{mathrsfs}
\usepackage{graphicx}
\usepackage{dcolumn}
\usepackage{bm}
\usepackage{color}

\usepackage{multirow}
\usepackage[colorlinks,bookmarks=false,citecolor=blue,linkcolor=red,urlcolor=blue]{hyperref}

\newcommand{\be}{\begin{equation}}
\newcommand{\ee}{\end{equation}}
\newcommand{\bea}{\begin{eqnarray}}
\newcommand{\eea}{\end{eqnarray}}

\definecolor{purple}{RGB}{128,0,128}

\begin{document}
\title{Robust topological superconductivity in spin-orbit coupled systems at higher-order van Hove filling}
\author{Xinloong Han}
\thanks{These authors contributed equally to the work.}
\affiliation{Kavli Institute for Theoretical Sciences, University of Chinese Academy of Sciences, Beijing 100190, China}

\author{Jun Zhan}
\thanks{These authors contributed equally to the work. }
\affiliation{Beijing National Laboratory for Condensed Matter Physics and Institute of Physics, Chinese Academy of Sciences, Beijing 100190, China}
\affiliation{School of Physical Sciences, University of Chinese Academy of Sciences, Beijing 100190, China}
\author{Fu-chun Zhang}
\affiliation{Kavli Institute for Theoretical Sciences, University of Chinese Academy of Sciences, Beijing 100190, China}
\author{Jiangping Hu}\email{jphu@iphy.ac.cn}
\affiliation{Beijing National Laboratory for Condensed Matter Physics and Institute of Physics, Chinese Academy of Sciences, Beijing 100190, China}
\affiliation{Kavli Institute for Theoretical Sciences, University of Chinese Academy of Sciences, Beijing 100190, China}
 \author{Xianxin Wu}\email{xxwu@itp.ac.cn}
 \affiliation{CAS Key Laboratory of Theoretical Physics, Institute of Theoretical Physics,
Chinese Academy of Sciences, Beijing 100190, China}

\date{\today}
\begin{abstract}

Van Hove singularities in proximity to the Fermi level promote electronic interactions and generate diverse competing instabilities.  It is also known that a nontrivial Berry phase derived from spin-orbit coupling can introduce an intriguing decoration into the interactions and thus alter correlated phenomena. However, it is unclear how and what type of new physics can emerge in a system featured by the
interplay between VHSs and the Berry phase. Here, based on a general Rashba model on the square lattice, we comprehensively explore such an interplay and its significant influence on the competing electronic instabilities by performing a parquet renormalization group analysis. Despite the existence of a variety of comparable fluctuations in the particle-particle and particle-hole channels associated with higher-order VHSs, we find that the chiral $p \pm ip$ pairings emerge as two stable fixed trajectories within the generic interaction parameter space, namely the system becomes a robust topological superconductor. The chiral pairings stem from the hopping interaction induced by the nontrivial Berry phase. The possible experimental realization and implications are  discussed. Our work sheds new light on the correlated states in quantum materials with strong SOC and offers fresh insights into the exploration of topological superconductivity.

\end{abstract}
 \maketitle
\noindent{\bf\emph{Keywords: Van Hove singularity, topological superconductors, Berry phase, Spin-orbital coupling}}

$\\$
\noindent{\bf 1. Introduction}
$\\$

Topological superconductivity has been attracting tremendous attention owing to the potential use in topological quantum computation of the hosted Majorana modes~\cite{Kane2010-RMP,SCZhang2011-RMP}. As the intrinsic topological superconductivity usually necessitates an exotic $p$-wave pairing, which is rare in nature, major experimental efforts focus on realizing synthetic $p$-wave states through exploiting the superconducting proximity effect in heterostructures~\cite{Lutchyn2018,OregY2010,LutchynRM2010,SauJD2010,DengMT2016,FuL2008,WangMX2012}, including nanowire-superconductor and topological-insulator-superconductor setups. In both cases, spin-orbit coupling, generating band splittings or topological surface states, plays an indispensable role in the synthetic pairing but has no direct interplay with electronic interactions.

In two-dimensional (2D) hetero-structures, interfaces and moir\'{e} materials, fascinating correlated phenomena often emerge~\cite{Reyren2007,GozarA2008,LiuCJ2021,QinSY2009,ZhangT2010,CaoY2018,CaoY2018-2,TangYH2020,Regan2020}, such as superconductivity. The ubiquitous spin-orbital coupling (SOC), whose strength can be tuned by external electric fields~\cite{Nitta1997,Miller2003}, is capable of  significantly modifying the electronic properties~\cite{Caviglia2010,Herranz2015}. Apart from that, another essential feature of SOC is that it introduces an intriguing Berry curvature field on the single-particle wavefunctions and thus a nontrivial geometric phase in the reciprocal space, which can dramatically alter interaction-driven phenomena~\cite{,Niu2006,Mao2011fwave-PRL,Qin2019}. On the other hand, van Hove (VH) singularities in the vicinity of Fermi energy in 2D materials, exhibiting divergent density of states (DOS), can generate diverse competing correlated states~\cite{VanHove1953,Schulz_1987,Furukawa1988,Chubukov2012,YaoH2015,Yuan2019,isobe2019supermetal}. For conventional VH singularities (VHSs), they feature logarithmically divergent DOS and particle-particle instability is usually dominant unless there is perfectly nested Fermi surface~\cite{Schulz_1987,Furukawa1988,Chubukov2012,YaoH2015}. In contrast, higher-order VHSs are characterized by stronger power-law divergent DOS and the corresponding fluctuations in particle-particle and particle-hole channels are comparable~\cite{shtyk2017electrons,Yuan2019,isobe2019supermetal,Efremov2019,classen2020competing,lin2020parquet}, resulting stronger competition. Once SOC is introduced in the VH dispersion, often occurring in 2D quantum materials, the concomitant band splitting lifts spin degeneracy and saddle points, moving away from high-symmetry momentum points, acquire a nontrivial Berry phase. They tend to put strong constraint on the electronic interactions and modify the competing instabilities. Therefore, it is desirable to explore the delicate interplay between electronic correlations associated with VHSs and geometric phases derived from SOC~\cite{Qin2019}.

In the present paper, we present a comprehensive study about the interplay between higher-order VHSs and the non-trival Berry phase through a weak-coupling renormalization group analysis. The non-interacting Hamiltonian is based on a general Rashba model on the square lattice, where higher-order VHSs are realized. Based on the simplified patch model, the parquet renormalization group calculations are adopted for two representative higher-order van Hove dispersion. We find that chiral $p \pm ip$ pairings emerge as two stable fixed trajectories in the interaction parameter space and are dominant with generic interaction setting, demonstrating a robust topological superconductivity. This chiral pairing is intimately correlated with a  pair hopping interaction involving a nontrivial Berry phase and the corresponding transition temperature is expected to be high. An interacting supermetal phase with power-law divergent density of states but no long-range order~\cite{isobe2019supermetal} can be stabilized at some parameter regime. The possible experimental realization and implications are also discussed.

$\\$
\noindent{\bf 2. Model and effective theory}\label{Model}
$\\$

\begin{figure}
	\includegraphics[width=0.45\textwidth]{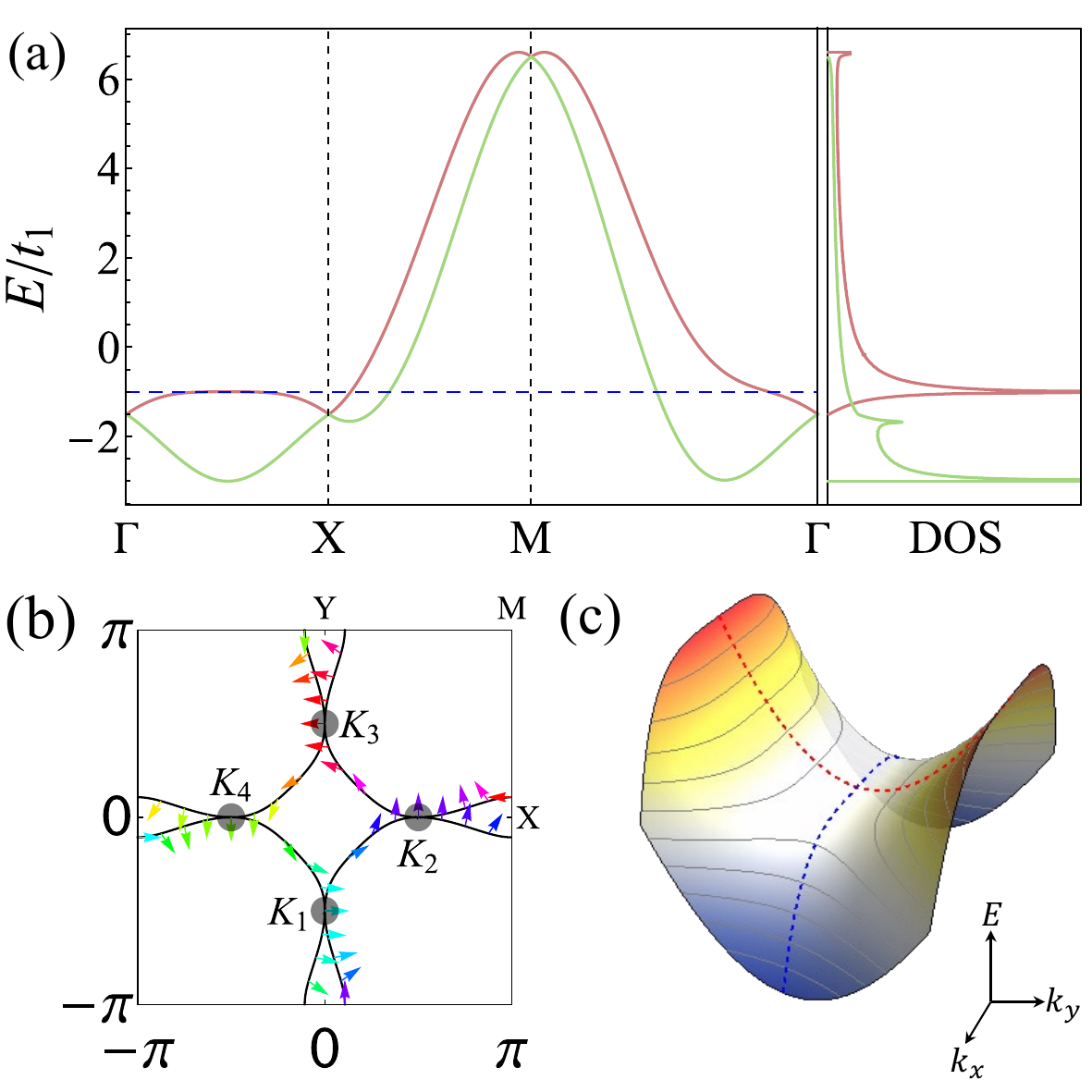}
	\caption{(a) Energy dispersion of the two Rashba-split bands and their density of states. The positive/negative helicity band is denoted by red/green line. Four higher-order VHSs are located at the Fermi level (blue dashed line) and the adopted parameters are $\lambda_R=t_1, ~t_2=-t_1/2,~t_3=-\lambda_R/8,~\mu_0=-t_1$. (b) Fermi surface of the positive-helicity band hosting higher-order VHSs and its spin texture  which is denoted by different colors based on their directions. (c) 2D dispersion around the higher-order saddle point with $E(\mathbf{k})=A(-k_x^4+k_y^2)$.}
	\label{fig:Fig1}
\end{figure}

We start with a generic tight-binding model with nearest-neighbor (NN) Rashba SOC terms on a square lattice,
\begin{equation}
	\begin{aligned}
	\mathcal{H}_{\text{TB}}=&-\sum_{\langle i, j\rangle_n \sigma} t_{n} (c_{i\sigma}^{\dagger} c_{j\sigma} + \text{h.c.})  -\mu_0\sum_{i\sigma}n_{i\sigma}  \\
	&+\frac{i}{2} \lambda_R \sum_{\langle ij\rangle,\sigma,\sigma^{\prime}} c_{i\sigma}^{\dagger}\left(\bm{\sigma}_{\sigma \sigma'} \times \mathbf{d}_{i j}\right)_z c_{j\sigma'},
	\end{aligned}
\end{equation}
where $c_{i\sigma}$ is the electron annihilation operator on the lattice site $i$ with a spin projection $\sigma= \uparrow,\downarrow$, $t_n$ is the hopping parameter between the $n$-th NN sites and $\lambda_R$ is the Rashba SOC constant between the NN sites with $\mathbf{d}_{i j}$ being the connecting vectors and $\bm{\sigma}$ being the Pauli matrices. Two conventional VHSs appear along the $X-\Gamma$ or $X-M$ lines in two helicity space upon including SOC. With hopping up to the next NN sites, the positive-helicity band $E_{+}(\mathbf{q})$ hosts a saddle point on the $\Gamma-X$ line and it is located at $P^+=(\pi-\arctan\frac{\lambda_R}{2t_1+4t_2},0)$. The corresponding low-energy dispersion expanded up to the quadratic terms reads $E_{P^+}(\mathbf{k})=\mu_{\text{VH}}-D_1 k^2_x + D_2 k^2_y$
 where $(k_x,k_y)$ is the deviation from the $P^+$ point, here $\mu_{\text{VH}}=-2t_1+D_0$, $D_1=D_0$ and $D_2=t_1+\frac{1}{2}D_0-4t_2(1+2t_2)/D_0$ with $D_0=\frac{1}{2}\sqrt{\lambda_R^2+4(t_1+2t_2)^2}$.
With further inclusion of longer-range hopping parameters, the analytical positions of VHSs cannot be obtained but our numerical calculations show that one coefficient of quadratic terms can vanish, realizing a higher-order VHS. In Fig. \ref{fig:Fig1}a, we display a representative tight-binding band dispersion with four higher-order VHSs in the vicinity of the Fermi level. Around the saddle point on the $\Gamma-X$ line, the low-energy dispersion is
$E(\mathbf{k}) = Ak_y^2-Bk_x^4$ with constants $A, B>0$
and it displays flat dispersion along the $k_x$ direction (as shown in Fig. \ref{fig:Fig1}a and c, contributing a power-law divergent density of states (DOS) $\nu(E)\propto |E|^{-\kappa}$ with an exponent $\kappa=1/4$. As shown in Fig. \ref{fig:Fig1}b, the corresponding Fermi surface is endowed with an intricate spin texture and this introduces an intriguing geometric phase for the four VHSs, where the Fermi surface touches tangentially (black circles in Fig. \ref{fig:Fig1}b). Other kinds of higher-order VHSs with distinct DOS exponents $\kappa$ can also be realized in this model and details are given the Supplementary materials. An interesting case is the VH dispersion
with $E(\mathbf{k}) =Ak_x^3-3Bk_xk_y^2$
and $\kappa=1/3$, where three Fermi surfaces touch at the saddle points.

\begin{figure}
\includegraphics[width=0.4\textwidth]{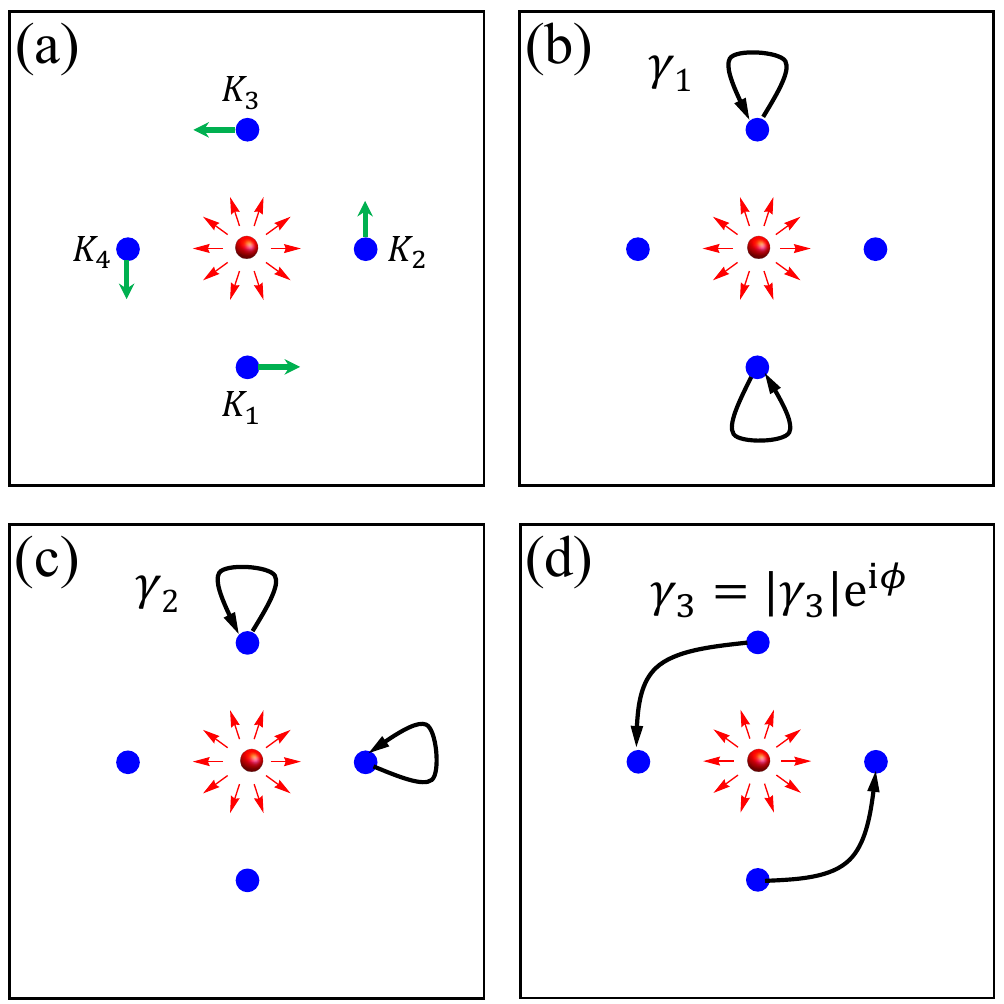}
\caption{(a) Low-energy patch model on the square lattice. The green arrow denotes the spin texture in each patch from the Dirac monopole (red sphere) at the center.   (b--d) Three allowed interactions from the momentum conservation and fermionic nature and the pairing hopping process ($\gamma_3$) obtains a nontrivial Berry phase. }
\label{fig:Fig2}
\end{figure}

We turn to study the correlation effect at the VH filling. Owing to the divergent DOS of VHSs, the collective electronic phenomena are expected to be dominated by the electronic scattering around these saddle points. Therefore, we can simplify the above non-interacting Hamiltonian and come to a low-energy patch model,
\begin{equation}
	\begin{aligned}
		\mathcal{H}_{\text{0}}=&\sum_{\alpha=1}^{4} \sum_{|\mathbf{k}|<k_\Lambda}\hat{c}_{\alpha\mathbf{k}}^{\dagger}[\epsilon_{\alpha}({\bf k})-\mu]c_{\alpha\mathbf{k}},
	\end{aligned}
\end{equation}
where $\mathbf{k}$ is centred at the saddle point $\alpha=K_i$ ($i=1,~2,~3,~4$), as illustrated in the Fig. \ref{fig:Fig2}a, with a patch size $k_\Lambda$ associated with the ultraviolet energy cutoff $\Lambda$. The $\epsilon_{\alpha}({\bf k})$ is the general low-energy dispersion around the higher-order saddle point $\alpha$. The non trivial spin texture of these VHSs is analogous to putting a monopole at the center of the Brillouin zone, as shown in Fig. \ref{fig:Fig2}a.
The nesting property and exponent $\kappa$ in the DOS are related to the VH dispersion and several kinds of higher-order VHSs have been discussed and realized in various platforms~\cite{shtyk2017electrons,Efremov2019,Yuan2019,isobe2019supermetal,classen2020competing}.
Our following treatment works for general VHSs and we choose two typical higher-order VH dispersion to show the essential characteristics:
(a) $\epsilon^e_{K_1}(\mathbf{k})= A(k_x^2-k_y^4)$ and two Fermi surface touch tangentially at the saddle point~\cite{Yuan2019,isobe2019supermetal},
(b) $\epsilon^o_{K_1}(\mathbf{k})= A(k_y^3-3k_xk_y^2)$ and three Fermi surfaces touch at the saddle point~\cite{shtyk2017electrons}.
The dispersion around other patches can be obtained $\epsilon_{ K_i}(\mathbf{k})=\epsilon_{K_1}(\hat{C}^{i-1}_4\mathbf{k})$ with $\hat{C}_4$ being the $\pi/2$ clockwise rotation operation.

Within this patch model, we consider all possible electron-electron interactions between the patches and only three interactions are allowed \cite{Furukawa1988,LeHur2009,Chubukov2012,Metzner2012},
\begin{eqnarray}
		&\mathcal{H}_{\text{int}}&=\frac{1}{2N}\sum_{\alpha=1}^{4}
		\mathop{\sum\nolimits^{\prime}}\limits_{|\mathbf{k}_1|,...,|\mathbf{k}_4|<k_\Lambda}
		\Big[\gamma_1
c^{\dagger}_{\bar{\alpha}\mathbf{k}_1}c_{\alpha\mathbf{k}_2}^{\dagger}c_{\alpha\mathbf{k}_3}c_{\bar{\alpha}\mathbf{k}_4}
		+\nonumber \\
		&&\gamma_2  c_{\langle\alpha\rangle\mathbf{k}_1}^{\dagger}  c^{\dagger}_{\alpha\mathbf{k}_2}
		 c_{\alpha\mathbf{k}_3}c_{\langle\alpha\rangle\mathbf{k}_4}
		+(|\gamma_3|e^{i\phi} c_{\bar{\alpha}+1\mathbf{k}_1}^{\dagger}c^{\dagger}_{\alpha+1\mathbf{k}_2} c_{\bar{\alpha}\mathbf{k}_3}\nonumber \\
		&& \times c_{\alpha\mathbf{k}_4}+h.c.)
		\Big],
\end{eqnarray}
where $\sum'_{|\bf{k}_1|,...,|\bf{k}_4|<k_\Lambda}=\sum_{|\mathbf{k}_1|,...,|\bf{k}_4|<k_\Lambda}\delta_{\bf{k}_1+\mathbf{k}_2,\mathbf{k}_3+\mathbf{k}_4}$ and $N$ is the number of unit cells. The patch $\langle \alpha\rangle$ and $\bar{\alpha}$ denotes the NN and the opposite patches of the patch $\alpha$, respectively.
As shown in Fig. \ref{fig:Fig2}b--d, $\gamma_1$ and $\gamma_2$ are the density-density interactions between opposite patches and NN patches while $\gamma_3$ is the pair hopping interaction. Remarkably, due to the presence of a Dirac monopole, this pair hopping process will accumulate a nontrivial phase as $\text{e}^{\text{i}\phi}$ with $\phi\in [-\pi,\pi ]$~\cite{swavewithSOC-PRL2008,Niu2006,Mao2011fwave-PRL,Qin2019}. In the tetragonal systems, this phase is fixed to $\phi=\pm \frac{\pi}{2}$ due to the four-fold rotational symmetry and fermionic nature. This even applies for the case with an out-plane magnetic field. In contrast to the hexagonal systems~\cite{Qin2019}, $\phi=0$ is not allowed in our case but corresponds to a vanishing $\gamma_3$ due to the lattice rotational symmetry and fermionic nature. This nontrivial Berry phase, as shown in the following, plays a pivotal role in determining the leading instability.

\begin{figure}[tp]
	\includegraphics[width=0.48\textwidth]{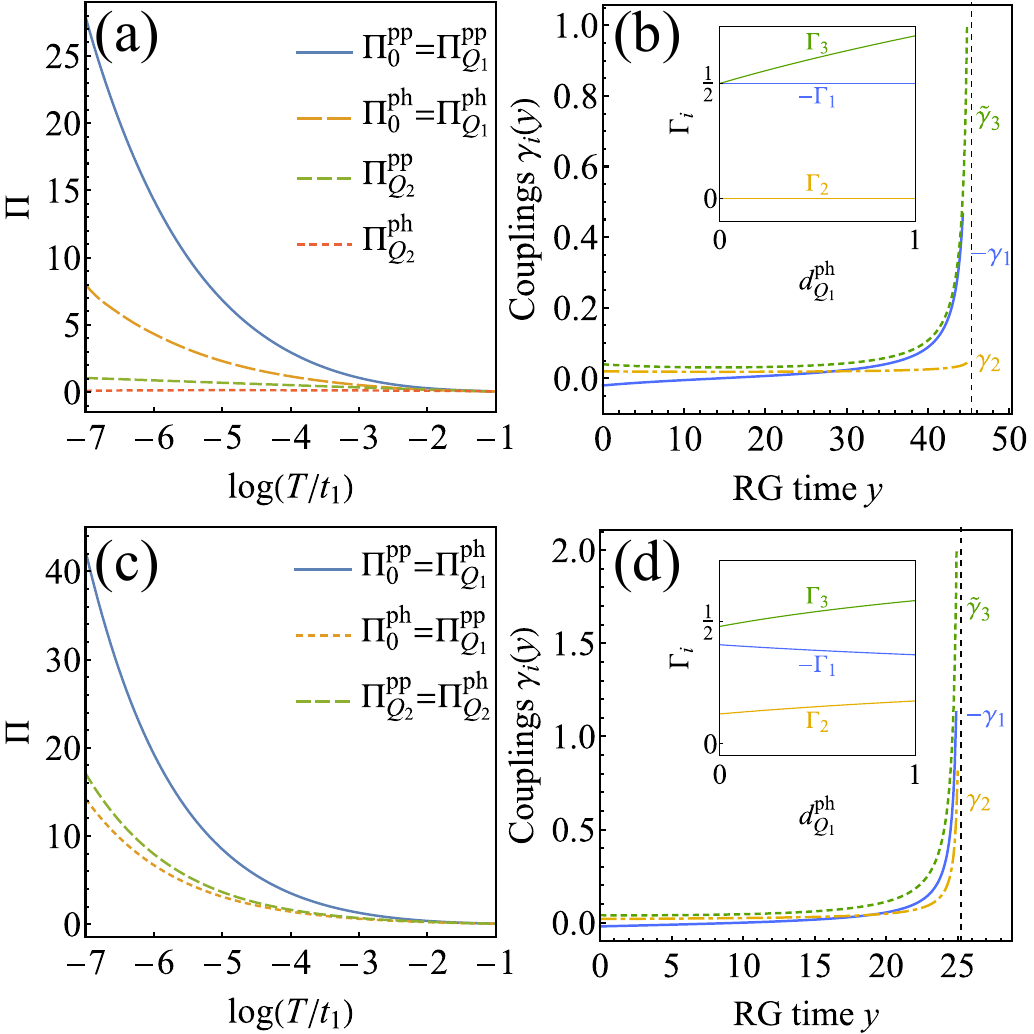}
	\caption{Susceptibilities (a,~c) as functions of $\log(T/t_1)$ and RG flows (b,~d) with initial repulsive interaction $\gamma_{1,2}(0)=0.02$ and $\tilde{\gamma}_3(0)=0.04$. Inset figures show the critical interactions $\Gamma_i$ as functions of $d^{\text{ph}}_{\mathbf{Q}_1}$ with $\tilde{\Gamma}_3=-i\Gamma_3$. The upper and bottom panels denote the cases of the higher-order VH dispersion with $\epsilon^e_{\alpha}(\mathbf{k})$ and $\epsilon^o_{\alpha}(\mathbf{k})$, respectively. In (a) $\Pi_{0}^{\text{pp,ph}}$ and $\Pi_{{\bf Q}_1}^{\text{pp,ph}}$ manifest power-law diverging, while $\Pi_{{\bf Q}_2}^{\text{pp,ph}}$ diverge as the logarithmic behavior. In (c) all particle-particle and particle-hole fluctuations behaves as a power-law diverging. In (b), we take nesting parameters as $d_{{\bf 0},{\bf Q}_1}^{\text{pp}}=1$, $d^{\text{ph}}_{{\bf 0},{\bf Q}_1}=1/4$ and $d_{{\bf Q}_2}^{\text{ph,pp}}=0$. In (d), we take $d_{{\bf 0},{\bf Q}_1}^{\text{pp}}=1$, $d^{\text{ph}}_{{\bf 0}}=d^{\text{pp}}_{{\bf Q}_1}=1/3$ and $d_{{\bf Q}_2}^{\text{ph,pp}}\simeq 0.39$.
	}
	\label{fig:Fig3}
\end{figure}

$\\$
\noindent{\bf 3. Susceptibilities and RG equations}
$\\$

Next, we perform renormalization group analysis to unveil competing electronic instabilities. Before deriving the RG equations, we first analyse particle-particle (pp) and particle-hole (ph) susceptibilities to determine which kinds of scattering channels are relevant near the Fermi energy. The static susceptibilities are defined as  $\Pi^{\text{pp/ph}}_{\mathbf{Q}}=\pm T\sum_{n}\int_{\mathbf{k}}G(\mathbf{k},\omega_n)G(\mp \mathbf{k}+ \mathbf{Q},\mp \omega_n)$. Here $G(\mathbf{k},\omega_n)=1/(i\omega_n-\xi_{\alpha,\mathbf{k}})$ is the bare Green's function of electrons with energy $\xi_{\alpha,\mathbf{k}}=\epsilon_{\alpha}(\mathbf{k})-\mu$ and fermionic Matsubara frequency $\omega_n=(2n+1)\pi T$. We focus on the VH filling $\mu=0$ and asymptotic limit $\Lambda \gg T$.
The particle-particle and particle-hole susceptibilities with zero momentum transfer in the patch model acquire the leading power-law divergence $\Pi^{\text{pp}}_{\mathbf{0}}=\frac{1}{\kappa} \Pi^{\text{ph}}_{\mathbf{0}} = C_0 T^{-\kappa}$ (more details see the Supplementary materials).

Two relevant momentum transfer are between opposite patches $\mathbf{Q}_1=2\mathbf{Q}_{\alpha}$ and NN patches $\mathbf{Q}_2=\mathbf{K}_{\alpha}-\mathbf{K}_{\alpha+1}$.
While, behaviours of susceptibilities at $\mathbf{Q}_1$ and $\mathbf{Q}_2$
strongly depend on the dispersion around the saddle point. By adopting numerical calculations, we plot all particle-particle and particle-hole susceptibilities for $\epsilon^e_{\alpha}(\mathbf{k})$ and $\epsilon^o_{\alpha}(\mathbf{k})$ as a function of the temperature in the Fig. \ref{fig:Fig3}a and c, respectively. In the former case, susceptibilities with $\mathbf{Q}_1$ in both channels are power-law divergent as $\Pi^{\text{pp/ph}}_{{\bf Q}_1}\propto 1/T^{1/4}$ due to $\epsilon^e_{\alpha}(\mathbf{k})=\epsilon^e_{\bar{\alpha}}(-\mathbf{k})$ and others with $\mathbf{Q}_2$ are constant or logarithmically divergent thus negligible. In the latter case, however, all susceptibilities with $\mathbf{Q}_1$ and $\mathbf{Q}_2$ are comparable and $\Pi^{\text{ph}/\text{pp}}_{\mathbf{Q}_1}=\Pi^{\text{pp}/\text{ph}}_{\mathbf{0}}\propto 1/T^{1/3}$ due to $\epsilon^o_{\alpha}(\mathbf{k})=-\epsilon^o_{\bar{\alpha}}(-\mathbf{k})$. To quantify the relative susceptibilities with respect to the Cooper instability, we introduce the nesting parameter,
\begin{equation}
	\begin{aligned}
		d^{\text{pp/ph}}_{\mathbf{q}}=\lim_{T/\Lambda\to 0} \partial\Pi^{\text{pp/ph}}_{\mathbf{q}}(T)/\partial\Pi^{\text{pp}}_{\mathbf{0}}(T).
	\end{aligned}
\end{equation}
Here a non-zero $d^{\text{pp/ph}}_{\mathbf{q}}$ represents a relevant susceptibility $\Pi^{\text{pp/ph}}_{\mathbf{q}}$. For the $\epsilon^e_{\alpha}(\mathbf{k})$, $d^{\text{pp}}_{\mathbf{0}}=d^{\text{pp}}_{\mathbf{Q}_1}=1$, $d^{\text{ph}}_{\mathbf{0}}=d^{\text{ph}}_{\mathbf{Q}_1}=1/4$, $d^{\text{pp}}_{\mathbf{Q}_2}=d^{\text{ph}}_{\mathbf{Q}_2}=0$.
For the $\epsilon^o_{\alpha}(\mathbf{k})$,
$d^{\text{pp}}_{\mathbf{0}}=d^{\text{ph}}_{\mathbf{Q}_1}=1$, $d^{\text{ph}}_{\mathbf{0}}=d^{\text{pp}}_{\mathbf{Q}_1}=1/3$, $d^{\text{pp}}_{\mathbf{Q}_2}=d^{\text{ph}}_{\mathbf{Q}_2}\simeq 0.39$ (more details see the Supplementary materials). It is also worth emphasizing that these nesting parameters depend on the low-energy dispersion of VHSs and their positions are irrelevant. The determination of these parameters for a general higher-order VH dispersion needs numerical calculations but the essential feature is that fluctuations in multiple pp and ph channels are comparable, in contrast to the conventional VHS~\cite{YaoH2015,Qin2019}.

By using the dominant susceptibilities as building blocks and neglecting self-energy corrections, we can derive RG equations for the dimensionless interactions $\bar{\gamma}_i=\gamma_i \partial \Pi^{\text{pp}}_0/\partial s$ including both tree-level and one-loop terms with the RG time $s=\ln{(\Lambda/T)}$. We find that all corresponding fixed points are generally unstable (see the Supplementary materials) and thus we focus on their strong-coupling trajectories, where certain interactions diverge towards a critical RG time. In this case, the tree-level terms are irrelevant and we can conveniently simplify the RG equations for the dimensional interactions~\cite{classen2020competing,lin2020parquet},
\begin{equation}\label{floweq}
	\begin{aligned}
		&\partial_y \gamma_1 =(d^{\text{ph}}_{\mathbf{Q}_1}-1)\gamma_1^2 -2 d^{\text{ph}}_{\mathbf{0}}\gamma_2^2-|\gamma_3|^2,\\
		&\partial_y \gamma_2 =(d^{\text{ph}}_{\mathbf{Q}_2}-d^{\text{pp}}_{\mathbf{Q}_2})\gamma_2^2 -2d^{\text{ph}}_{\mathbf{0}}\gamma_1\gamma_2+d^{\text{ph}}_{\mathbf{Q}_2}|\gamma_3|^2,\\
		&\partial_y \gamma_3 =-2 d^{\text{pp}}_{\mathbf{0}} \gamma_1\gamma_3 +4d^{\text{ph}}_{\mathbf{Q}_2} \gamma_2\gamma_3,
	\end{aligned}
\end{equation}
where we change the RG time to $y=\Pi^{\text{pp}}_{\mathbf{0}}(T)-\Pi^{\text{pp}}_{\mathbf{0}}(\Lambda)$. As $\gamma_1$ and $\gamma_2$ terms conserves
the number of electrons at each patch separately but $\gamma_3$ does not, the $\beta$ function of $\gamma_3$ must contain an overall factor of itself. Further considering the absence of $\gamma^*_3$ terms in the $\beta$ function, the associated phase of $\gamma_3$ remains fixed during the RG evolution. Since the $\beta$ function for $\gamma_1$ becomes negative definite, i.e., $\partial_y \gamma_1 <0$, $\gamma_1$ always tends to flow negative infinity from a negative initial value, promoting instabilities in both pp and ph channels.

To determine the possible competing instabilities, we introduce the test vertices in various pp and ph channels,
$\delta \mathcal{H}_{\text{I}}=\sum_{\alpha\alpha'}[\Delta^{\text{I}}_{\alpha\alpha'}{c}^{\dagger}_{\alpha}{c}^{(\dagger)}_{\alpha'}+h.c.]$. Under the RG flow, they obtain one-loop corrections through the divergent susceptibilities, $\partial_{y} \Delta^{\text{I}}=- d^{\text{I}}\mathcal{M}^{\text{I}}\Delta^{\text{I}}$ with $d^{\text{I}}=d^{\text{pp/ph}}_{\mathbf{q}}$ and $\mathcal{M}^{\text{I}}$ being the corresponding driving interaction matrix. These orders can be further decoupled into different irreducible channels according to the lattice symmetry.
Approaching the critical RG time $y_c$, interactions diverge and usually flow into some fixed trajectories, which can be parameterized as $\gamma_i(y)=\Gamma_i/(y_c-y)$. Hence, the vertices will diverge as $\Delta_I(y)\propto (y_c-y)^{\eta_I}$ and the corresponding susceptibility scales as $\chi_{I} \propto (y_c-y)^{2\eta_{I}+1}$, implying that only vertices with $\eta_I<-1/2$ are relevant instabilities and the order with most negative $\eta_I$ is the leading instability. We list all possible orders in various channels and the corresponding exponents in the Table \ref{tab1} (for details see the Supplementary materials). The lifted spin degeneracy imposes strong constraints on the instabilities. In the particle-particle channel, the intra-patch pairing density wave (PDW) is forbidden and only the $p$-wave pairing between the opposite patches is allowed. The $p+ip$ and $p-ip$ pairing states get split owing to the pair hopping scattering $\gamma_3$ involving a Berry phase.

\begin{table}[t]
	\begin{ruledtabular}
		\caption{Divergent exponents $\eta_{\text{I}}$ for all competing orders in the pp and ph channels.}\label{tab1}
		\begin{tabular}{lllll}
			Particle-particle order  & Particle-hole order \\\hline
			$\eta^{\text{SC}}_{p\pm ip}=2d^{\text{pp}}_{\mathbf{0}}(\Gamma_1 \mp \tilde{
			\Gamma}_3)$ & $\eta^{\text{Pom}}_{s/d}=2d^{\text{ph}}_{\mathbf{0}}(\Gamma_1 \pm 2\Gamma_2)$ \\
			 & $\eta^{\text{Pom}}_{p}=-2d^{\text{ph}}_{\mathbf{0}}\Gamma_1$ \\
			 & $\eta^{\text{CDW}}_{\mathbf{Q}_1}=-2d^{\text{ph}}_{\mathbf{Q}_1}\Gamma_1$ \\
			$\eta^{\text{PDW}}_{\mathbf{Q}_2}=2d^{\text{pp}}_{\mathbf{Q}_2}\Gamma_2$ & $\eta^{\text{CDW}}_{\mathbf{Q}_2,\pm}=2d^{\text{ph}}_{\mathbf{Q}_2}(-\Gamma_2 \pm \tilde{
			\Gamma}_3)$ \\
		\end{tabular}
	\end{ruledtabular}
\end{table}

$\\$
\noindent{\bf 4. Competing instabilities}
$\\$

\begin{figure}[tp]
	\includegraphics[width=0.45\textwidth]{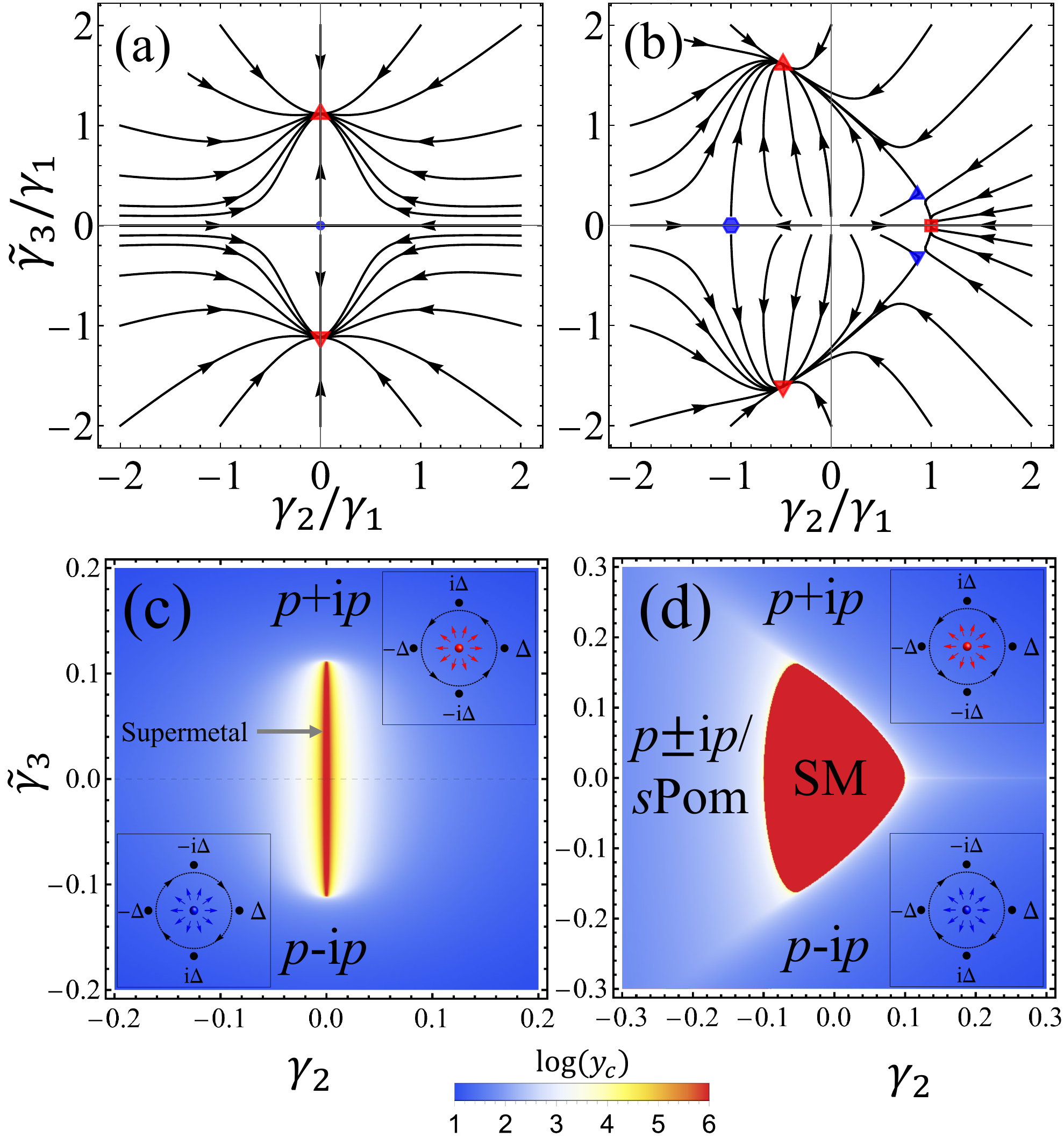}
	\caption{RG flow diagrams of the reduced RG equation \ref{rRG:eq1} in the parameter space $(\gamma_2/\gamma_1,\tilde{\gamma}_3/\gamma_1)$ for the $\epsilon_{\alpha}^e({\bf k})$ (a) and $\epsilon^o_{\alpha}({\bf k})$ (b), with $\gamma_1<0$. Phase diagrams for the higher-order VHSs with $\epsilon_{\alpha}^e({\bf k})$ (c) and $\epsilon^o_{\alpha}({\bf k})$ (d) for the initial interaction $\gamma_1(0)=0.1$. The color map represents the critical RG scale $\text{log}(y_c)$ and the red region, namely too large $y_c$, denotes that there is no instability. In (a,c), we take nesting parameters as $d_{{\bf 0},{\bf Q}_1}^{\text{pp}}=1$, $d^{\text{ph}}_{{\bf 0},{\bf Q}_1}=1/4$ and $d_{{\bf Q}_2}^{\text{ph,pp}}=0$. In (b,d), we take $d_{{\bf 0},{\bf Q}_1}^{\text{pp}}=1$, $d^{\text{ph}}_{{\bf 0}}=d^{\text{pp}}_{{\bf Q}_1}=1/3$ and $d_{{\bf Q}_2}^{\text{ph,pp}}\simeq 0.39$.
	}
	\label{fig:Fig4}
\end{figure}

We first study the characteristic flow with initial repulsive interactions.
With defining $\tilde{\gamma}_3=-i\gamma_3$, we plot the RG flows of the coupling constants from starting repulsive interactions as $\gamma_{1,2}(0)=0.02$ and $\gamma_3(0)=0.04i$ for higher-order VH fillings with $\epsilon_{\alpha}^e(\mathbf{k})$ and $\epsilon_{\alpha}^o(\mathbf{k})$ in the Fig. \ref{fig:Fig3}b and d, respectively. All interactions diverge approaching a critical RG time and the critical value in the latter case is smaller due to the additional boost of $\Pi^{\text{pp/ph}}_{\mathbf{Q}_2}$ related terms. These interactions flow into one fixed trajectory and the evolution of $\Gamma_i$ as function of $d^{\text{ph}}_{\mathbf{Q}_1}$ are displayed in the insets. We find that $\Gamma_2$ is zero for $\epsilon_{\alpha}^e(\mathbf{k})$, indicating that $\gamma_2$ diverges slower than $1/(y_c-y)$. For $\epsilon_{\alpha}^o(\mathbf{k})$, all three interactions are comparable. In this fixed trajectory, the dominant instability is the $p_x+ip_y$ pairing according to Table \ref{tab1}. To determine all fixed trajectories and their stability \cite{Chubukov2012,Herbut2007,Yang2010-PRB}, we introduce the following reduced RG equations for the relative interactions $x_i=\gamma_i/\gamma_1(i=2,3)$ with RG time proxy $\gamma_{1}$,

\begin{equation}\label{rRG:eq1}
	\begin{aligned}
		&\frac{\text{d} x_2}{d\ln |\gamma_1|} =\frac{(d^{\text{ph}}_{\mathbf{Q}_2}-d^{\text{pp}}_{\mathbf{Q}_2})x_2^2 -2d^{\text{ph}}_{\mathbf{0}}x_2+d^{\text{ph}}_{\mathbf{Q}_2}|x_3|^2}
		{d^{\text{ph}}_{\mathbf{Q}_1}-d^{\text{pp}}_{\mathbf{0}} -2 d^{\text{ph}}_{\mathbf{0}}x_2^2-d^{\text{pp}}_{\mathbf{0}}|x_3|^2} - x_2,\\
		&\frac{\text{d} x_3}{d\ln |\gamma_1|} =\frac{-2 d^{\text{pp}}_{\mathbf{Q}_1} x_3 +4d^{\text{ph}}_{\mathbf{Q}_2} x_2x_3}
		{d^{\text{ph}}_{\mathbf{Q}_1}-d^{\text{pp}}_{\mathbf{0}} -2 d^{\text{ph}}_{\mathbf{0}}x_2^2-d^{\text{pp}}_{\mathbf{0}}|x_3|^2} - x_3.
	\end{aligned}
\end{equation}
The nontrivial fixed trajectories $(\Gamma_i)$ of Eq. (\ref{floweq}) then will have a correspondence with the fixed points in Eq.(\ref{rRG:eq1}) as $(\Gamma_2/\Gamma_1,\Gamma_3/\Gamma_1)$. In the Fig. \ref{fig:Fig4}a and b,  we \textcolor{red}{first} illustrate the RG flow diagrams of ratios $x_{2,3}$ for higher-order VHSs with $\epsilon_{\alpha}^e(\mathbf{k})$ and $\epsilon_{\alpha}^o(\mathbf{k})$ with initial negative $\gamma_1(0)$ because of its negative defined of its RG flow equation shown in Eq. (\ref{floweq}). The common feature in both cases is that there are two stable fixed points at the positive and negative finite $\tilde{\gamma}_3/\gamma_1$ values, corresponding to the chiral $p_x\mp ip_y$ superconducting states (red point-up/point-down triangle). In contrast to the only two nontrivial fixed points for $\epsilon_{\alpha}^e(\mathbf{k})$, there are four more fixed points for the $\epsilon_{\alpha}^o(\mathbf{k})$. Three of them are unstable, corresponding to the $d$-wave Pomeranchuk order (blue hexagon) and additional chiral $p\pm ip$ pairing (blue triangles), and the rest one is stable (red square), which is a degenerate point for the $p\pm ip$ pairing and s-wave charge Pomeranchuk order (sPom). The degeneracy and stability are related to the dispersion and these states can be split once $\kappa$ deviates from $\frac{1}{3}$. We further display the phase diagrams for the repulsive initial interaction $\gamma_1(0)$ as depicted in Fig. \ref{fig:Fig4}c and d. It is apparent that the chiral superconductivity dominates the phase diagram in both cases. An interacting supermetal phase, exhibiting divergent DOS and finite dimensionless interaction but no long-range order~\cite{isobe2019supermetal}, emerges on certain line for $\epsilon_{\alpha}^e(\mathbf{k})$ and certain region for $\epsilon_{\alpha}^o(\mathbf{k})$ and there is a strong CDW fluctuation in the latter case (see the Supplementary materials). The three-fold degenerate state, corresponding to the fixed point (red square) in Fig. \ref{fig:Fig4}b, can appear for an initial strong attractive $\gamma_2$.

$\\$
\noindent{\bf 5. Discussions}
$\\$

In our calculations, the superconductivity pairing is dominantly from the positive-helicity band with large DOS, while the negative-helicity band crossing the Fermi level carries small DOS and obtains superconductivity by the proximity effect. For the realistic on-site Hubbard interaction $U$ and nearest neighbor interaction $V$, the coupling constants can be expressed as $\lambda_{1,2}(0)\simeq U+8V$, and $\lambda_3(0)\simeq U+4V(\cos(P^+_{x}))$ where $P^{+}$ denotes the position of VHSs (more details see the Supplementary materials Note2). Therefore we can tune the amplitude or even change its sign of $\lambda_3$ by moving the position of VHSs, although arising from repulsive $U$ and $V$. Hence by tuning the position of VHSs, it is possible to realize the transition between the $p+ip$ and $p-ip$ chiral superconducting states. This will result a nonzero Chern number in the superconducting state.

Despite comparable fluctuations in the multiple pp and ph channels at the higher-order VH filling, we find robust instability in the pp channel with zero momentum, determined by the negative definite interaction $\gamma_1$. The pair hopping scattering $\gamma_3$ involving a nontrivial Berry phase further promotes a chiral pairing in the band space, corresponding to a mixture between equal-spin and spin-singlet pairing due to the broken inversion symmetry. Owing to the power-law divergent DOS, the superconducting transition temperature scales as $T_c\propto \lambda^{1/\kappa}$ with $\lambda$ being coupling constant, compared with the case of conventional VHS $T_c\propto e^{-1/\sqrt{N_F\lambda}}$ where $N_F$ is the density state near the Fermi level. In the realistic systems, several kinds of higher-order VHSs has been realized in stoichiometric materials, heterostructures and twisted systems~\cite{2021arXiv210504546M,Efremov2019,Yuan2019,WuXX2019,BiZ2021,Guerci2022}. The crucial factor is to introduce an additional SOC and this can be achieved in inversion-breaking materials with heavy atom substitution. In heterostructures and moir\'{e} materials SOC can be further enhanced and tuned by the external electric field~\cite{BiZ2021,Guerci2022,PhysRevResearch.2.033087}. Once higher-order VHSs with SOC are realized, the electric gating approach can be used to introduce carrier doping and tune electronic instabilities. An important feature of chiral superconductivity is that it can host Majorana zero modes in the vortex, which can be detected using experimental technique, such as scanning tunneling microscope~\cite{WangDF2018}.

$\\$
\noindent{\bf 6. Conclusions and remarks}
$\\$

In summary, we show that the interplay of higher-order VHSs and a nontrivial Berry phase has a dramatic impact on the competing electronic instabilities within the RG formalism. The chiral $p\pm ip$ pairings, directly related to the pair hopping process involving a Berry phase, emerge as two stable fixed trajectories in the interaction parameter space and are dominant in the phase diagram with generic interaction setting. Our work demonstrates the robust chiral superconductivity in spin-orbit coupled systems at higher-order VH filling and offers fresh insights into the exploration of topological superconductivity. In contrast to the square lattice, the hexagonal system can host two more higher-order VHSs and the associated lattice symmetry is distinct, leading to a complex landscape of competing instabilities, which we leave for our future study.

$\\$
\noindent{\bf Conflict of interest}
$\\$

The authors declare that they have no conflict of interest.

$\\$
\noindent{\bf Acknowledgments}
$\\$

We acknowledge the supports by the Ministry of Science and Technology  (2022YFA1403901), the National Natural Science Foundation of China (11920101005,  11888101, and 12047503) and the New Cornerstone Investigator Program. Fu-Chun Zhang is partially supported by the Priority Program of Chinese Academy of Sciences (JZHKYPT-2021-08) and is also partially supported by Chinese Academy of Sciences  under contract No. JZHKYPT-2021-08.
Xinloong Han also acknowledges the supports from China Postdoctoral Science Foundation Fellowship (2022M723112).

$\\$
\noindent{\bf Author contributions}
$\\$

Xinloong Han, Xianxin Wu, and Jiangping Hu conceived this project. Xinloong Han, Jun Zhan, and Xianxin Wu performed the analytical and numerical calculations. Xinloong Han, Jun Zhan, Fu-Chun Zhang, Xianxin Wu, and Jiangping Hu jointly performed the analysis and wrote the paper.

\bibliographystyle{apsrev4-1}

\begin{thebibliography}{99}

\bibitem{Kane2010-RMP} Hasan M. Z. and Kane C. L. Colloquium: topological insulators. Rev Mod Phys 2010;82:3045-3067.

\bibitem{SCZhang2011-RMP} Qi X.L. and Zhang S.C. Topological insulators and superconductors. Rev Mod Phys 2011;83:1057-1110.

\bibitem{Lutchyn2018} Lutchyn R. M., Bakkers E. P. A. M., Kouwenhoven L. P., et~al. Majorana zero modes in superconductor-semiconductor heterostructures. Nat Rev Mater 2018;3:52-68.

\bibitem{OregY2010} Oreg Y., Refael G. and von Oppen F. Helical liquids and Majorana bound states in quantum wires. Phys Rev Lett 2010;105:077001.

\bibitem{LutchynRM2010} Lutchyn R. M., Sau J. D., and Das Sarma S. Majorana fermions and a topological phase transition in semiconductor-superconductor heterostructures. Phys Rev Lett 2010;105:077001.

\bibitem{SauJD2010} Sau J. D., Lutchyn R. M., Tewari S., et~al. Generic new platform for topological quantum computation using semiconductor heterostructures. Phys Rev Lett 2010;104:040502.

\bibitem{DengMT2016} Deng M. T., Vaitiekenas S., Hansen E. B., et~al. Majorana bound state in a coupled quantum-dot hybrid-nanowire system. Science 2016;354:1557-562.

\bibitem{FuL2008} Fu L. and Kane C. L. Superconducting proximity effect and Majorana fermions at the surface of a topological insulator. Phys Rev Lett 2008;100:096407.

\bibitem{WangMX2012} Wang M.-X., Liu C., Xu J.-P., et~al. The coexistence of superconductivity and topological order in the Bi$_2$Se$_3$ thin films. Science 2012;336:52-55.

\bibitem{Reyren2007} Reyren N., Thiel S., Caviglia A. D., et~al. Superconducting interfaces between insulating oxides. Science 2007;317:1196-1199.

\bibitem{GozarA2008} Gozar A., Logvenov G., Kourkoutis L. F., et~al. High-temperature interface superconductivity between metallic and insulating copper oxides. Nature 2008;455:782-785.

\bibitem{LiuCJ2021} Liu C., Yan X., Jin D., et~al. Two-dimensional superconductivity and anisotropic transport at KTaO$_3$ (111) interfaces. Science 2021;371:716-721.

\bibitem{QinSY2009} Qin S., Kim J., Niu Q., et~al. Superconductivity at the two-dimensional limit. Science 2009;324:1314-1317.

\bibitem{ZhangT2010} Zhang T., Cheng P., Li W.-J., et~al. Superconductivity in one-atomic-layer metal films grown on Si(111). Nat Phys 2010;6:104-108.

\bibitem{CaoY2018} Cao Y., Fatemi V., Demir A., et~al. Correlated insulator behaviour at half-filling in magic-angle graphene superlattices. Nature 2018;556:80-84.

\bibitem{CaoY2018-2} Cao Y., Fatemi V., Fang S., et~al. Unconventional superconductivity in magic-angle graphene superlattices.  Nature 2018;556:43-50.

\bibitem{TangYH2020}  Tang Y., Li L., Li T., et~al. Simulation of Hubbard model physics in WSe$_2$/WS$_2$ moiré superlattices. Nature 2020;579:353-358.

\bibitem{Regan2020}  Regan E. C., Wang D., Jin C., et~al. Mott and generalized Wigner crystal states in WSe$_2$/WS$_2$ moiré superlattices. Nature 2020;579:359-363.

\bibitem{Nitta1997} Nitta J., Akazaki T., Takayanagi H., et~al. Gate control of spin-orbit interaction in an inverted I${\mathrm{n}}_{0.53}$G${\mathrm{a}}_{0.47}$As/I${\mathrm{n}}_{0.52}$A${\mathrm{l}}_{0.48}$As heterostructure. Phys Rev Lett 1997;78:1335-1338.

\bibitem{Miller2003} Miller J. B., Zumbuhl D. M., Marcus C. M., et~al. Gate-controlled spin-orbit quantum interference effects in lateral transport. Phys Rev Lett 2003;90:076807.

\bibitem{Caviglia2010} Caviglia A. D. , Gabay M., Gariglio S., et~al. Tunable Rashba spin-orbit interaction at oxide interfaces.  Phys Rev Lett 2010; 104:126803.

\bibitem{Herranz2015} Herranz G., Singh G., Bergeal N., et~al. Engineering two-dimensional superconductivity and Rashba spin–orbit coupling in LaAlO$_3$/SrTiO$_3$ quantum wells by selective orbital occupancy. Nat Commun 2015;6:6028.

\bibitem{Niu2006} J. Shi and Q. Niu. Attractive electron-electron interaction induced by geometric phase in a Bloch band. Sci China Phys Mech Astron 2020;63:227422.

\bibitem{Mao2011fwave-PRL} Mao L., Shi J., Niu Q., et~al. Superconducting phase with a chiral $f$-wave pairing symmetry and Majorana fermions induced in a hole-doped semiconductor.  Phys Rev Lett 2011;106:157003.

\bibitem{Qin2019}  Qin W., Li L., and Zhang Z. Chiral topological superconductivity arising from the interplay of geometric phase and electron correlation. Nat Phys 2019;15:796-802.

\bibitem{VanHove1953} Van Hove L. The occurrence of singularities in the elastic frequency distribution of a crystal Phys Rev 1953;89:1189-1193.

\bibitem{Schulz_1987} Schulz H. J. Superconductivity and antiferromagnetism in the two-dimensional Hubbard model: Scaling Theory. Europhysics Letters 1987;4:609.

\bibitem{Furukawa1988}  Furukawa N., Rice T. M., and  Salmhofer M. Truncation of a two-dimensional Fermi surface due to quasiparticle gap formation at the saddle points.  Phys Rev Lett 1998;81:3195.

\bibitem{Chubukov2012} Nandkishore R., Levitov L. S., and Chubukov A. V. Chiral superconductivity from repulsive interactions in doped graphene. Nat Phys 2012;8:158-163.

\bibitem{YaoH2015} Yao H. and Yang F. Topological odd-parity superconductivity at type-II two-dimensional van Hove singularities. Phys Rev B 2015;92:035132.

\bibitem{Yuan2019} Yuan N. F. Q., Isobe H., and Fu L. Magic of high-order van Hove singularity. Nat Commun 2019;10:5769.

\bibitem{isobe2019supermetal}  Isobe H. and Fu L. Supermetal. Phys Rev Res 2019;1:033206.

\bibitem{shtyk2017electrons} Shtyk A., Goldstein G., and Chamon C. Electrons at the monkey saddle: A multicritical Lifshitz point. Phys Rev B 2017;95:035137.

\bibitem{Efremov2019} Efremov D. V., Shtyk A., Rost A. W., et~al. Multicritical Fermi surface topological transitions. Phys Rev Lett 2019;123:207202.

\bibitem{classen2020competing} Classen L., Chubukov A. V., Honerkamp C., et~al. Competing orders at higher-order van Hove points. Phys Rev B 2020;102:125141.

\bibitem{lin2020parquet}  Lin Y.-P. and  Nandkishore R. M. Parquet renormalization group analysis of weak-coupling instabilities with multiple high-order van Hove points inside the Brillouin zone.  Phys Rev B  2020;102:245122.


\bibitem{LeHur2009}  Hur K. L. and  Maurice Rice T. Superconductivity close to the Mott state: From condensed-matter systems to superfluidity in optical lattices. Annals of Physics 2009;324:1452.

\bibitem{Metzner2012}  Metzner W., Salmhofer M., Honerkamp C., et~al. Functional renormalization group approach to correlated fermion systems. Rev Mod Phys 2012;84:299-352.

\bibitem{swavewithSOC-PRL2008}  Zhang C., Tewari  S., R. M. Lutchyn, et~al. ${p}_{x}+i{p}_{y}$ superfluid from $s$-wave interactions of fermionic cold atoms. Phys Rev Lett 2008;101:160401.

\bibitem{Herbut2007}  Herbut I. A Modern approach to critical phenomena. (Cambridge University Press, 2007).

\bibitem{Yang2010-PRB}  Vafek O. and  Yang K. Many-body instability of coulomb interacting bilayer graphene: Renormalization group approach.  Phys Rev B 2010;81:041401(R).

\bibitem{2021arXiv210504546M} Markiewicz R. S., Singh B., Lane C., et~al. High-order van Hove singularities in cuprates and related high-Tc superconductors. arXiv 2021;2105:04546.

\bibitem{WuXX2019}  Di Sante D.,  Wu X.,  Fink M., et~al. Triplet superconductivity in the Dirac semimetal germanene on a substrate. Phys Rev B 2019;99:201106(R).

\bibitem{BiZ2021}  Bi Z. and Fu L. Excitonic density wave and spin-valley superfluid in bilayer transition metal dichalcogenide. Nat Commun 2021;12:642.

\bibitem{Guerci2022}  Guerci D., Simon P., and  Mora C. Higher-order van Hove singularity in magic-angle twisted trilayer graphene.  Phys Rev Res 2022;4:L012013.

\bibitem{PhysRevResearch.2.033087}  Pan H., Wu F., and Das Sarma S. Band topology, Hubbard model, Heisenberg model, and Dzyaloshinskii-Moriya interaction in twisted bilayer ${\mathrm{WSe}}_{2}$. Phys Rev Res 2020;2:
033087.

\bibitem{WangDF2018} Wang D.,  Kong L.,  Fan P., et~al. Evidence for Majorana bound states in an iron-based superconductor. Science 2018;362:333-335.
\end{thebibliography}

\end{document}